# Bringing the Internet to Schools: US and EU policies

Dr Michelle S. Kosmidis

European Commission

## Abstract

*The Internet is changing rapidly the way people around the world communicate, learn, and work. Yet the tremendous benefits of the Internet are not shared equally by all. One way to close the gap of the 'digital divide' is to ensure Internet access to all schools from an early age. While both the USA and EU have embraced the promotion of Internet access to schools, the two have decided to finance it differently. This paper shows that the main costs of Internet access to schools are not communications-related (telecommunications and Internet services) but rather non-communications-related (hardware, educational training, software). This paper goes on to discuss whether the identified costs should be financed in any way by the universal service obligations funded by the telecommunications industry/sector/consumers (sector specific) or a general governmental budget (educational budget).*

Acknowledgements: I would like to thank Mark Scanlan, Don Stockdale, and Howard Williams for their comments. The author takes full responsibility for any errors. The views expressed are strictly personal and are not necessarily shared by the Organisation to which the author belongs.
Prepared for delivery at the 2001 TPRC Meeting, October 27-29, 2001, Alexandria, VA, USA (TPRC-2001-076).

## 1. Introduction

The Internet has changed the way people communicate, learn, and work. Efforts have been made to integrate information technology and Internet into the educational systems for a number of compelling reasons. The new technology is considered as an important pedagogic tool, which complements rather than replaces traditional teaching at schools. It prepares the young workforce to be more employable in the new economy, which is more information-based. Furthermore, such integration could help to eliminate the 'digital divide' through providing unequivocally access to all students to new technologies regardless of social or economic background.

Measures have been taken to promote Internet access and usage in elementary and secondary schools in both the United States (US) and European Union (EU).[1] At a first stage public policy appears to support Internet access to schools (i.e., Internet access to a computer laboratory of each school), while at a second stage it may concentrate on Internet access specifically to classrooms. This distinction is significant as the latter policy choice (Internet access to each classroom) involves huge costs.

It is beyond the scope of this paper to discuss the precise effects of information technology on students' learning, which is analysed in other research studies.[2] Instead it focuses on regulatory and economic issues related to the costs of effective Internet access to schools: communications costs (telephone and Internet Service Provider (ISP) charges); and non-communications costs, such as costs for hardware (computers) and internal connections (such as network wiring, modems, routers, network file services, wireless local area networks (LANs)), maintenance and technical support, teachers' training, and specialised software.

The question this paper addresses is who should pay for any of the aforementioned costs. To tackle this question, available data on costs of Internet access to schools will be analysed in order to identify the type of costs related to such access, the magnitude of communications costs, and the proportion of these costs in relation to the total information communication technology (ICT)[3] school expenditures. The key principles for covering such costs should be operator neutrality, technological neutrality (no type of technology should be favoured), and should not distort competition.

This paper shows that communications costs comprise only a small proportion of ICT school expenditures. Empirical evidence in both the EU and US suggests that competition gives incentives for operators to provide discounted tariffs for Internet access to schools. The issue is whether there should be any additional support, and how it should be funded.

---

[1] The development of Internet in schools can vary from an information awareness tool for students (a few PCs for each school); information resource for staff with a few PCs in the staff room; learning computer applications as a subject in laboratories with approximately 30 terminals; and as a platform for teaching with a number of PCs in each classroom. This paper focuses on the last two options.

[2] A number of studies show the positive effects of technology in schools and classrooms on student achievement, motivation level, and speed of learning. See summary on pages 22-23 of U.S. Department of Education (2000a); also PIC (1997), pp.34-40.

[3] ICT school expenditures include communications and non-communications costs as defined earlier.

This paper finds that non-communications costs are significantly higher than the cost of communications services. The EU Member States finance the additional costs through general national budgets or joint public/private partnerships, while the US imposes a 'tax' specifically on telecommunications operators providing interstate and international services. It concludes that telecommunications operators should not be obliged to pay for these non-communications costs. As we will see later, such taxation on the telecommunications industry is costly, inefficient, and not necessarily equitable.

The paper also examines the difficult question of the effectiveness of each policy approach by analysing the penetration rates of Internet access to schools in the United States and the 15 EU Member States. Indicators like number of pupils per computer (with Internet access), and number of instructional rooms with computers (with Internet access) demonstrate the use of Internet as a learning tool. Furthermore, the type of connection and speed is crucial for the use of Internet in schools. Data show that the US is ahead of the EU average – although some Member States (like Sweden and Denmark) are doing better than the European average and have comparable results with the United States. Data also suggest that the US already had a high penetration rate of Internet access to schools before the adoption of the E-Rate program in 1998 raising the question of whether full access would have been achieved anyway without such program.

This paper will describe the US policy where the universal service fund supports Internet access to eligible schools (section 2). It will present the EU policy of encouraging operators to provide special (flat rate) tariffs (at the dial-up layer) to schools and/or funding through a general government budget (section 3). Section 4 compares policy outcomes showing the penetration of Internet access to schools in the US and EU. In conclusion, this paper compares the two policy lines for Internet access to schools.

## 2. Internet access to schools in the USA

*Introduction*

Prior to 1984, the availability of affordable telephone access to all American households was achieved through AT&T's internal pricing structure. After the 1984 AT&T break-up, and as competition was being introduced into the long-distance market, an explicit Universal Service Fund was established in order to sustain affordable access for high cost rural areas and low income users.

In 1994, the Clinton-Gore National Information Infrastructure (NII) initiative announced its commitment to develop a seamless national network of information and telecommunications services. One goal was to connect all K-12 schools and instructional rooms (e.g. every classroom, computer labs, and library/media centers) to the Internet by the year 2000.

Consistent with this initiative, the Telecommunications Act of 1996 expanded the concept of universal service obligations to include support for schools, libraries and rural health providers in the 1996 Telecommunications Act. The federal government's choice of supporting Internet access to schools through telecommunications and not educational policy may be a result of its limited policy instruments and budgetary resources in the latter area, which is largely dominated by the state level.

In implementing the 1996 Act, the FCC, in May 1997, outlined a plan in its Universal Service Order to guarantee affordable access to telecommunications services for all eligible schools (and libraries[4]).[5] The federal universal service costs for Internet access to schools and libraries (the so-called E-Rate program) was capped at $2.25 billion per year[6] out of the total annual explicit universal service costs of $4.65 billion. Funding from the E-Rate program was available starting from January 1, 1998.

**Table 1**
**Annual explicit federal universal service fund**

|  | Caps on annual collection $ Billion |
|---|---|
| High cost areas and small telephone companies[1] | 1.6 |
| Low income users[2] | 0.4 |
| Schools and libraries | 2.25 |
| Rural healthcare | 0.4 |
| **Total explicit universal service costs** | **4.65** |

Notes:
1. Includes Loop costs, Local Switching support (LSS), Long Term Support (LTS). This item is mainly concerned with alleviating any negative consequences of tariff averaging.
2. Includes Lifeline Assistance, Link-up program.

The following sections describe the selection criteria for schools receiving discounts, and evaluate the actual allocation of funding of the E-Rate program.

*Administrative process for allocating funds*

This section describes the administrative structure for allocating funds, which is quite complex and leads to burdensome and inefficient processes in receiving funds. The School and Libraries Division (SLD) of the Universal Service Administrative Company (USAC), a non-profit entity established by the FCC, allocated universal service support for schools and libraries.

The criteria for determining eligible schools, level of discounts, and eligible services are as follows. Eligible schools are those (public or non-public) elementary and secondary schools that do not operate as a for-profit business, and do not have an endowment exceeding $50 million.[7] Eligible schools must develop a state-approved educational technology plan in order to show how they intend to integrate the use of technologies into their curricula, and to show that there is sufficient budget to acquire non-discounted elements of plan (i.e., hardware, software).

---

[4] For consistent comparison with the EU this paper does not discuss in detail Internet access to libraries. For the record, however, libraries received 4% of total E-Rate funding during the first two years of the program.

[5] At the same time, the FCC reformed the interstate access charge system in order to adjust to a competitive market.

[6] This amount is matched with equivalent state funding.

The level of discounts (ranging from 20 to 90%) received by eligible schools (and libraries) depends on the poverty level measured as the percentage of students eligible for the subsidised National School Lunch Program,[8] and the rural location of the school. A matrix has been established for the link between school's eligibility and the level of discount. For instance, schools or consortiums that have more than 75% of their students receiving subsidised school lunches receive a 90% discount on eligible services.

Services eligible for discount include specified telecommunications services,[9] Internet access services (email service), and internal connections.[10] The fund will not finance customer premises equipment (CPE) or hardware, such as personal computers.[11] Where requests exceed available funds, telecommunications and Internet services have priority over internal connections. Schools and libraries have the maximum flexibility to purchase the services and technologies to suit their needs (wireline, wireless or cable technology), making it technology neutral.

Once the application for discount has been approved, the school will receive the applicable discount on its telecommunications services, Internet access and/or internal connections, and pay the remaining portion of the costs to the vendor or service providers. The vendor or service providers present the discounted bills, and get reimbursed by the Fund Administrator (USAC).

In order to fund the various universal service programs, all telecommunications carriers providing end-user interstate and international telecommunications services are obliged to contribute a fee to the fund based on their interstate and international end-user telecommunications revenues.[12]

---

[7] Eligible schools and libraries can create a consortium with sufficient demand to attract competitors and negotiate lower rates. There are four types of applicants: libraries (or library consortia), schools, school districts, and SLC consortia.

[8] This program is administered by the U.S. Department of Agriculture and state agencies that supply free or inexpensive lunches to economically disadvantaged students. Libraries use the discount percentage of the school district in which they are located.

[9] Local, long-distance and toll charges, dedicated lines, leased DS-1, T-1, ISDN, xDSL, Directory assistance charges.

[10] Telecommunications wiring, routers, hubs, network file servers, switches, hinds, network servers, certain networking software, wireless LANs, Private Branch Exchange (PBX), installation and basic maintenance of internal connections.

[11] The fund will not finance CPE and hardware, such as personal computers (with the exception of network file servers), telephone handsets, fax machines, modems, nor will it finance asbestos removal, the costs of tearing down walls to install wiring, repairing carpets, repainting, learning software, or teacher/librarian training.

[12] Interstate telecommunications includes, but is not limited to: 'cellular telephone and paging services; mobile radio services; operator services; PCS; access to inter-exchange service; special access; wide area telephone services (WATS); toll-free services; 900 services; MTS; private line; telex; telegraph; video services; satellite services; and resale services' in FCC (1997) *Universal Service Order*.

*Policy debates and evaluation of actual allocation of funding*

During the introduction of the E-Rate program there was much debate concerning the scope and fairness of the program. This section focuses on the main concerns over the new funding scheme, and evaluates the actual allocation of funding.

One concern was over the extension of the scope of universal service to finance Internet access to schools by taxing the telecommunications industry. Indeed, in 1997 three companies (GTE, BellSouth, and SBC Communications) appealed the FCC's Universal Service Order claiming that the E-Rate was an illegal tax.[13] They argued that a federal agency such as the FCC did not have the authority to impose a tax on them -- only Congress did. Furthermore, they argued that it was unfair to burden the telecommunications industry with network infrastructure costs of schools

Indeed when looking at the allocation of the total funding in terms of type of services during January 1998 - June 2000, the highest proportion of the funding is allocated to internal connections. More specifically, on average during this period 58% of the E-Rate funds financed the acquisition of equipment and services for internal building connections, 34% telecommunications services (which also includes dedicated lines), and 8% Internet access costs.

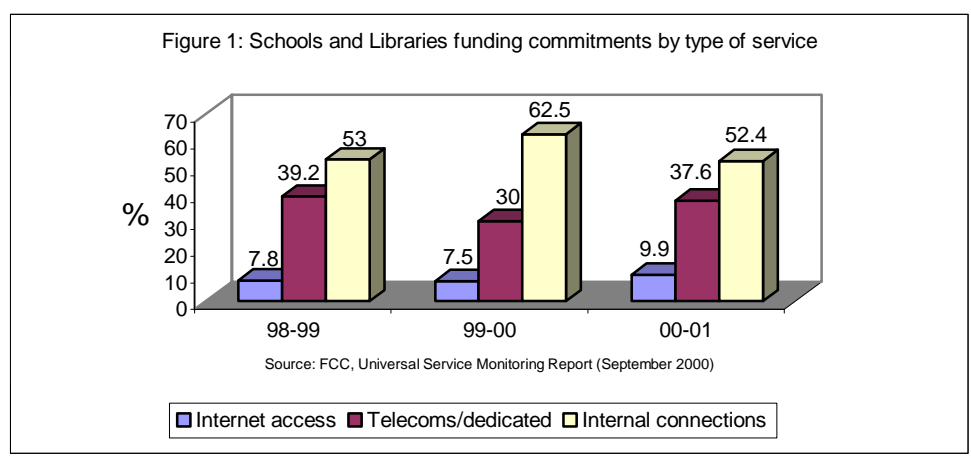

Figure 1: Schools and Libraries funding commitments by type of service

Source: FCC, Universal Service Monitoring Report (September 2000)

Notes:
1. First cycle was January 1, 1998-June 30, 1999; the next July 1, 1999- June 30, 2000. Subsequent years run from July 1 to June 30 of the following calendar year.
2.  00-01: For funds committed between July 1- August 2000.

Another concern was the lack of operator neutrality in the FCC's choice of who should contribute to the E-Rate fund. Only interstate telecommunications carriers, and not Internet Service Providers (ISPs), had to contribute to the fund. There appears to be a policy inconsistency in the E-Rate program where ISPs are exempt from contributing to a service that is entirely Internet-related.

---

[13] Eventually BellSouth and SBC withdrew their appeal, and only GTE remained as a litigant. In July 1999 the Court of Appeals rejected GTE's complaint, and in May 2000, the US Supreme Court refused to hear the litigants' appeal, therefore terminating E-Rate opposition. See US Department of Education (2000b), pp. 19-21.

In July 1998, the three largest long-distance telecommunications companies (AT&T, MCI and Sprint) started expressly passing on the cost to customers with a 'universal service' line item on their phone bills. This started a debate over the impact of the E-Rate program on consumers, as a response, the FCC scaled back its annual finance from $2.25 to $1.9 billion to cover a longer period of 18 months (from January 1, 1998 - June 30, 1999) instead of the initially planned 12 months.[14]

Recent studies (Prieger 1998; WIK 2000) show that the federal universal service obligations (USO) tax on the telecommunications industry is actually quite costly and inefficient. Prieger (1998) estimated a deadweight loss or 'inefficiency' of end-user revenue taxation ranging between $1.2 - 4 billion per year. This revenue taxation has a negative impact on consumer behaviour as it is directly passed on to them causing a reduction of services consumed, and consequently a reduction of industry revenues.

Since one of the goals of the E-Rate program was to reduce the digital divide in the US, it is important to evaluate the funding allocation to the poorest school districts. According to a report by the U.S. Department of Education (2000b) only 14% of total E-Rate funds went to the highest poverty public school districts (with more than 75% of their students eligible for reduced price lunches) during the period of January 1998-June 2000. Application rates are lower in poorest school districts because they do not have the human and financial resources to apply for funds, they cannot afford the remaining 10% of charges after the E-Rate discount, or lack basic infrastructure (like electrical outlets) and computers to incorporate the new technology. E-Rate funding per student in particular for internal connections was especially high in these higher poverty districts.

In evaluating the procedures of allocating E-Rate funding, the General Accounting Office[15] issued a report in 2001, which showed that about 24% of the funds committed for years 1998-1999 remained unspent. Slow bureaucratic application and invoice procedures, and endless paperwork have been blamed for this failure. It usually takes several months until USAC sends an approval letter to the schools, and that is when everything slows down or stops. Schools at that stage do not claim the funding they are entitled to.

There is an ongoing debate on whether to transfer the federal E-Rate program to the states to be distributed through a block grant program rather than the current burdensome application process.[16]

**3. Internet access to schools in the EU**

In Europe there is a growing consensus that its citizens will have to learn to use the new information and communications tools from an early age if they want to succeed in a knowledge-based society and labour market. Internet access to European schools is becoming more common although there is a wide variation of such access among EU Member States with Denmark and Sweden being the most advanced.

---

[14] The FCC also wanted to prevent the E-Rate charges from rising faster than access charge reductions (in July 1997 the FCC reduced access charges by the amount of $1.9 billion).

[15] GAO is the investigative arm of Congress, which examines the use of public funds, and evaluates federal programs and activities.

[16] The current Bush Administration discussed the possibility of removing the administration and oversight of the E-Rate program from the FCC to the Department of Education.

The EU Action Plan *Learning in the Information Society (1996-1998)* was launched in 1996 to encourage various activities to connect schools to communications networks. It was only three years later in December 1999 that the European Commission launched the *eEurope* Initiative with the goal of bringing Europe on-line.[17] At the Lisbon Summit in March 2000, the Heads of State and Government committed themselves to a number of measures, including target dates, to bring *eEurope* forward. One of *eEurope* targets is to connect all schools to the Internet by the end of 2001, and all classrooms to high-speed Internet and multimedia resources by the end of 2002 – two years later than the US 2000 target. Some Member States already had action plans, projects, and goals to connect schools earlier.

In contrast with the US regime, the current (and the recently proposed) EU legislative framework does not allow any subsidy associated to Internet access to schools and usage to be included in the universal service obligations[18] funding schemes. Instead the EU policy choice is to connect schools to the Internet either by encouraging competition among operators to provide 'special tariffs' or service packages to schools and/or through joint public/private initiatives. Actual data from the United Kingdom show that communications costs amount to a small proportion of the total ICT costs.

*Competition with special tariff packages*

The European Commission encourages National Regulatory Authorities (NRAs) to allow incumbent and new operators to offer *special tariffs* to schools (at the dial-up layer) for Internet access and usage.[19] These special tariffs should not distort competition and incumbent operators should not abuse their dominant position, for instance, through predatory pricing. Operators will be willing to compete for the provision of such service to schools as it is hoped that it will get students hooked up to Internet, which will have a spill-over effect on usage at homes.

In Europe telecommunications charges are typically usage dependent (contrary to the USA) and therefore unpredictable and difficult to handle in annual school budgets. The benefits of 'special packages or tariffs' offered to schools are that they are by far a *less expensive* and *more predictable cost* compared to estimated 'standard' rates.[20] At times special flat rate prices for schools are one fifth of estimated 'standard' prices. For instance, British Telecom offers annual access to schools with unlimited usage for

---

[17] See http://europa.eu.int/information_society/eeurope/action_plan/index_en.htm.

[18] The scope of universal service obligations include the provision of voice telephony, fax and dial up internet connection, as well as public pay phones, emergency call access, operator and directory services. It must be noted that during the ongoing adoption of the new EU regulatory framework there was discussion to include Internet access to schools in the scope of universal service obligations.

[19] In the current regulatory framework, the Voice Telephony Directive (98/10/EC) allows for the provision of special tariff schemes, and requires transparency and cost orientation for tariffs of carriers possessing Significant Market Power (SMP).

[20] All estimated 'standard' calculations in this section include an annual PSTN rental cost, a usage cost (based on an assumed use of 200 hour per month for each school - therefore a total of 50 hours per week per school), and annual ISP charges. Tariffs from Teligen (2000).

$590 (euro 631) with a Public Services Telephone Network (PSTN) connection, and for $1047 (euro 1,120) with an Integrated Services Digital Network (ISDN) connection. This would compare to estimated UK standard rate of $7760 with PSTN, and $4030 with ISDN access in year 2000. In Portugal, Telepac provides 200 hours of Internet access per month to schools with PSTN for an annual fee of $186 (euro 199), but local charges apply in addition.

**Table 2**
**Special packages for Internet access to schools in EU Member States (Jan. 2000)[1]**

|     | Company | PSTN [2] Annual charge (dollars $) | ISDN [2] Annual charge (dollars $) | Comments |
|---|---|---|---|---|
| **D** | T-Online | | | Free ISDN and Internet access to schools. No telephone charges for access to Internet. |
| **E** | Telefónica | 56 | 95 (Basic) | Since 1998, a 3 year agreement between Telefónica and Madrid's local government of free access and reduced usage tariffs to schools in the region. Unlimited access and usage. |
| **F** | Wanadoo | 70 (incl. VAT) | | Launched in 1998, flat annual charge per single connection for unlimited Internet access. Telephone usage charges are billed separately. |
| **F** | Cegetel & Havas | 120 | | Unlimited Interest access to schools; associated telephone charges are bundled at $693 (euro 741) per year, giving 380 hours of usage. |
| **IRL** | Telecom Eireann | | | Free multi-media PC. Telephone line free of rental charges for 2 years. Free internet access for 2 years. Telephone usage credit of up to 183 hours per year for 2 years. |
| **IT** | TIN | 1st 18 mths: free 2nd 18 mths: 128 After annual:128 (excl. VAT) | 1st 18 mths: free 2nd 18 mths: 314 After annual:314 (excl. VAT) | Unlimited Internet access. Telephone usage charges billed separately. |
| **AUT** | Netway | | | Free Internet access for 80 hours per month to all Austrian schools. Schools liable for telephone usage charges. |
| **P** | Telepac | 186 (incl. VAT) | | 200 hours of Internet access per month (over 200 hours suspended service). Local rate charges apply in addition. |
| **UK** | BT | 590 (excl. VAT) | 1047 (excl. VAT) | Internet access and usage between 8:00-18:00. |
| **UK** | Cable companies | | | UK cable companies charge $1.9 (euro 1.6) per pupil per year for unlimited access, and including all telephone calls to the Internet. |

Note:
1. This table includes some non-comparable and non-exhaustive data of Internet access offers to schools as of January 2000 (Source:Teligen, 2000). Table uses annual 1999 exchange rate for euro 1= $1.07 (Eurostat).
2. PSTN allows for one connection; Basic ISDN (bandwidth of 128 kbit/s) allows for two simultaneous connections and Primary ISDN allows for 32 simultaneous connections.

The following table presents the estimated EU average 'standard' price for Internet access to schools, which is considerably higher than 'special' tariffs. Such standard prices vary widely according to each Member State.

**Table 3**
**Estimated EU average 'standard' price for Internet access to schools (Jan. 2000)** [1,2,3]

|                  | Per school | Per pupil | Comments                      |
|------------------|------------|-----------|-------------------------------|
| **PSTN**         | $3902      | $3902     | One user per time             |
| **ISDN (BASIC)** | $4020      | $2010     | Assumes two simultaneous users|
| **ISDN (PRIMARY)**| $7327     | $229      | 32 simultaneous users         |
| **Leased Line 2 MBIT** | $11275 | $352     | 32 simultaneous users         |

Note:
1. All estimated 'standard' calculations in this section include an annual PSTN rental cost, a usage cost (based on an assumed use of 200 hour per month for each school - therefore a total of 50 hours per week per school), and annual ISP charges. Tariffs from Teligen (2000).
2. PSTN allows for one connection; Basic ISDN (bandwidth of 128 kbit/s) allows for two simultaneous connections and Primary ISDN allows for 32 simultaneous connections.
3. There are considerable differences between Member States with estimated 'standard' prices for Internet access to schools varying between $3263-$15.058 for ISDN-Primary connection, and $6075-$20.530 for leased line (2 Mbit) connection.

Competition issues in relation to abuse of dominant position and competition distortion have been raised through the provision of 'special' or free Internet access to schools. In accordance with Article 82 (ex Article 86) of the Treaty a dominant operator cannot 'directly or indirectly impose unfair purchase or selling prices or other unfair trading conditions.' In the United Kingdom[21] and France[22] the incumbent operators provided very low price Internet access to schools. Competitors complained that they could not compete with the incumbent's offer since they had to pay a high interconnection rate.[23] In both countries these cases were resolved by enforcing lower interconnection rates allowing the new entrants to compete and in some cases even provide cheaper rates to schools (such as the case of cable operators in the United Kingdom).

*Joint public/private initiatives for bringing Internet to schools*

EU Member States are free to finance ICT expenses from general budgets. These initiatives can be supported either by public (national, regional and local level) sector alone, or by joint public/private partnerships. The choice of private industry participation must be in accordance to EU transparency and non-discrimination principles. Government initiatives are invariably focused on Internet access, hardware, network infrastructure, and/or teacher training.

---

[21] In May 1997 British Telecom (BT) was the first operator in Europe, which proposed to offer flat rate PSTN and Basic ISDN tariffs for Internet access to schools. Following a public consultation, in October 1997, OFTEL established the cost floors for the prices that BT could charge schools in order to make sure that other operators could compete for this service. For more details see OFTEL (1997a & b).

[22] In 1998 a group of new operators (Cegetel, Bouygues, and Colt Telecom) filed a complaint against France Telecom (FT) for providing discounted fees for Internet access to schools. FT planned to fix prices at $950 a year for Internet access for 10 hours a day for 10 PCs. New operators could not compete with this offer because they had to pay high interconnection charges to FT on top of their costs. After negotiations, FT agreed to lower its interconnection rates.

[23] In Europe, competitive operators that serve ISPs have to pay an interconnection fee to incumbent operators. On the contrary, in the United States, under the reciprocal compensation scheme set forth in the 1996 Act, CLECs that serve ISPs actually receive termination fees from the ILEC.

An example of public/private partnership is Germany's 'Schulen ans Netz' (Schools Online). The German Federal Ministry of Education and Research (BMBF) and Deutsche Telekom (DT) jointly launched a federal with the goal to connect 10,000 schools to Internet between 1996 and mid-1999 ($33 million). The two partners strengthened their commitment with a further $55 million to connect all 36,000 schools to Internet by the end of 2001 (therefore a total of around $88 million). The federal states are providing additional funding, while there are numerous other sponsors from industry.

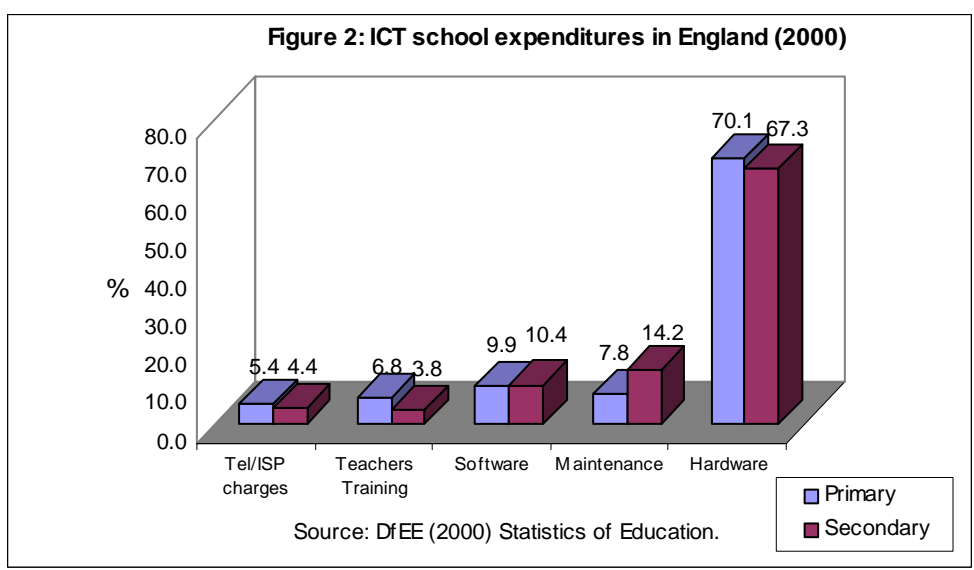

Data on the progress of UK governmental initiative (National Grid for Learning (NGfL)) show that communications costs are very low in relation to the total ICT school expenditures. A recent survey shows the total annual ICT expenditures for teaching and learning in primary and secondary schools was around $440 million in 2000 (DfEE 2000).[24] Hardware accounted for the highest cost of around 70% of the total ICT expenditures.[25] Communications costs (telecommunications and ISP) accounted only for 4-5% or around $20 million of the total ICT expenditures.[26] These communications charges are estimated at $599 (£370) per primary school and $2965 (£1830) per secondary school.

---

[24] More specifically, total ICT school expenditure for primary schools was $199 mn (£123 mn) and for secondary schools was $240mn (£148 mn). The average expenditure per school was $11016 (£6800) for primary schools, and $ 67716 (£41,800) for secondary schools. The average expenditure per pupil was $49 (£30) in primary schools and $76 (£47) in secondary schools. (DfEE, 2000). Annual 1999 exchange rate £1=$1.62 (Bank of England).

[25] Hardware includes computers, peripheral equipment, upgrades and replacements. A US research by RAND (1996) and California Department of Education (1996) also shows a similar high portion of ICT expenditures on hardware. See PIC (1997). Furthermore, the amount spent on internal wiring, which is the biggest expense in the US, is not clearly specified in the survey.

[26] A US study also supports that telecommunications costs account for 4-11% of total technology expenditures. PIC (1997).

After analysing the different policy choices in financing Internet access to schools in the United States and the EU, the following section will analyse and compare the outcome of each policy choice.

## 4. Comparing policy outcomes in the US and EU Member States

A recent US survey shows that public schools in the United States have nearly reached the goal of connecting every school to the Internet. The percentage of public schools connected to the Internet has increased from 35% in 1994 to 98% in 2000. The successful policy outcome in the US may not necessarily be a result of effective policy choice of funding Internet access to schools through the E-Rate program in the US. First, one must note that there was significant (and even higher) annual growth of penetration before the disbursal of the E-Rate program in November 1998. Second, Internet access to all schools was already at 89% in 1998 (Crandall and Waverman, 2000). Furthermore, the aforementioned in section 2 GAO findings (which show that around 25% of the 1998-1999 E-Rate funds still remain unused) put into question the actual impact of the E-Rate program.

**Table 4**
**Percentage of schools with Internet access in the USA (1994-2000)**

| School characteristics | 1994 | 1995 | 1996 | 1997 | 1998 | 1999 | 2000 |
|---|---|---|---|---|---|---|---|
| Elementary | 30 | 46 | 61 | 75 | 88 | 94 | 97 |
| Secondary | 49 | 65 | 77 | 89 | 94 | 98 | 100 |
| **All public schools** | 35 | 50 | 65 | 78 | 89 | 95 | 98 |
| **All instructional rooms** | 3 | 8 | 14 | 27 | 51 | 64 | 77 |

Source: National Center for Education Statistics (2001): *Internet Access in US Public Schools and Classrooms: 1994-2000.* May 2001.

The EU average of schools connected to the Internet was lower than the US at 89% at the beginning of 2001. Some Member States like Sweden and Denmark, which financed earlier in the mid-1990s Internet connection to schools through their state educational budget, have slightly better results than the United States.[27]

Internet connection to classrooms reveals the policy aim to use the new technology as a pedagogical tool. In the US, the percentage of public school instructional rooms with Internet access increased dramatically from 3% in 1994 to 77% in 2000. However, only 60% of classrooms in schools with the highest penetration of students in poverty (75% or more students eligible for the free lunch program) are connected to the Internet. In Europe there are no solid data on this aspect reflecting for the moment its current priority to connect to schools rather than classrooms.

Other important indicators are the ratio of students per PC, and students per PC with Internet access. In the USA in 2000, the ratio of students per instructional computer in public schools was 5 to 1 and per instructional computer with Internet access was 7 to 1 (NCES 2001). The average EU ratio of students per PC was 12 to 1, and per Internet PC was 24 to 1 in early 2001 (CEC 20001). Again, when looking at ratio of pupils per PC in secondary schools in particular Member States, Denmark, Sweden, and Finland are as advanced as the USA compared to the rest of the EU countries surveyed (for instance in Portugal the ratio is 18 pupils per computer).

---

[27] Although one must note that the EU data is from early 2001.

**Table 5**
**Internet connection to schools in the EU* (and USA) (2000-01)**

|  | Number of pupils per computer | | % of schools connected to Internet | |
| --- | --- | --- | --- | --- |
|  | Primary | Secondary | Primary | Secondary |
| **Belgium** | 11 | 8 | 90 | 96 |
| **Denmark** | 4 | 1 | 98 | 99 |
| **Germany** | 23 | 14 | 90 | 98 |
| **Greece** | 67 | 17 | 22 | 58 |
| **Spain** | 14 | 14 | 91 | 95 |
| **France** | 16 | 10 | 63 | 97 |
| **Ireland** | 12 | 8 | 96 | 99 |
| **Italy** | 22 | 9 | 87 | 98 |
| **Luxembourg** | 2 | 6 | 86 | 100 |
| **The Netherlands** | 8 | 9 | 91 | 100 |
| **Austria** | 11 | 9 | 53 | 95 |
| **Portugal** | 26 | 18 | 56 | 91 |
| **Finland** | 7 | 7 | 99 | 99 |
| **Sweden** | 10 | 4 | 100 | 100 |
| **The United Kingdom** | 12 | 6 | 93 | 98 |
| **EU average*** | 12 | | 89 | |
| **The United States \*\*** | 5 | | 98 | |

Source: CEC (2001); Eurobarometer surveys, February 2001. * EU average includes all schools (primary, secondary and professional/schools). **NCES, 2001: US data from 2000.

The type of network connection and speed determines the ability of the school to connect to the Internet. In the US, 77% of schools used dedicated leased lines, 11% dial-up connections, and 24% other types of connections (including ISDN) in 2000. But even before the introduction of the E-Rate program there were some evident improvements in network connections to public schools.

**Table 6**

**Percentage of US public schools with Internet access using different type of connections (1996-00)**

| School characteristics | 1996 | 1997 | 1998 | 1999 | 2000 |
| --- | --- | --- | --- | --- | --- |
| Dedicated line | 39 | 45 | 65 | 72 | 77 |
| Dial-up connection | 74 | 50 | 22 | 15 | 11 |
| Other types |  |  | 26 | 23 | 24 |

Sources: National Center for Education Statistics (2001). *Internet Access in US Public Schools and Classrooms: 1994-2000.* May 2001.

In Europe, in February 2001, Internet connection to schools is dominated by narrowband technologies (such as dial-up connections, and ISDN). Around 72% of schools use ISDN connections, 33% standard dial-up connections, while only 5% use ADSL, and 6 % use cable modems (multiple connections are possible) (CEC 2001). The high proportion of narrowband connectivity suggests that the use of Internet in schools is less developed in Europe compared to the US. Furthermore, the limited use of leased lines in Europe may be a result of the existing expensive tariffs.

## 5. Conclusions

Governments in the US and EU Member States seem to agree that Internet access to schools is important for the 21$^{st}$ century. The issue of controversy is who should pay for the costs of Internet access to schools. An important issue emerges on a 'digital divide' among schools.

This paper shows that communications usage costs do not serve as the most important barrier to the development of Internet access to schools in the US and the EU. Empirical evidence in the US and UK show that communications costs (telecommunications and Internet access) account for a small portion of the total school ICT expenditure. Available data also show that the primary factor affecting ICT school costs is the purchase and installation of computers and other hardware. Secondary costs relate to maintenance, teachers' training, and software.

Telecommunications operators should not be obliged to pay for costs related to infrastructure and educational purposes. Imposing extra financial obligations on telecommunications operators may undermine the development of competition (infrastructure and services) for telecommunications, and does not necessarily reach equity objectives.

Looking at a different question of the effectiveness of policy outcome, the USA is well ahead of the EU average in bringing Internet to schools. High penetration rates may simply be a result of early market liberalisation and Internet take-off in the United States. Progress may also be related to the fact that Internet usage in the United States is not linked to per minute usage, as in Europe. Notably the more liberalised countries like Sweden and Denmark have similar results with the USA. Furthermore, it is not clear whether the E-Rate program (which was disbursed in November 1998) had a major impact on Internet access to schools. Considering that Europe started much later than the US, the penetration rate is increasing rapidly through special usage tariffs and government educational initiatives.

At times the comparison of policy outcomes between the US and EU was difficult due to the lack of data in the EU Member States. The *e-Europe Initiative* and its related 'Benchmarking' exercise should be more rigorous and systematic in gathering more relevant data on Internet access to classrooms, type and age available PCs, and general ICT school expenditures. Collection of other socio-economic data of schools would also be useful to evaluate the policy outcome, most notably to assess the possibility of a 'digital divide' between European schools.

In developing policy to address Internet access and usage at schools, the general principles should be that policy is operator neutral, technology neutral, and does not distort competition. Specific non-communications expenditures related to assisting schools should come from education or national budgets and not by introducing additional distortions to the telecommunications sector.

**References**

CEC (2001). *eEurope 2002 Benchmarking: European youth into the digital age, forthcoming.*

Crandall, R.W. and Leonard Waverman (eds.) (2000). *Who pays for universal service? When telephone policies become transparent.* The Brookings Institution: Washington, DC.

DfEE (2000). *Statistics of Education: Survey of Information and Communications Technology in School.* Issue No 07/00, October 2000.

Federal Communications Commission (FCC) (1997). *Universal Service Order*, FCC Docket 96-45.

Federal Ministry of Education and Research & Federal Ministry of Economics and Technology (1999). *Innovation and Jobs in the Information Society of the 21$^{st}$ Century: Action Programme by the German Government.* Druck and Verlag, Bonn.

GAO (2001). *Schools and Libraries Program: Update on E-Rate Funding.* GAO-01-672, Washington, D.C.

National Agency for Education (2000). *Schools and Computers 1999 – a quantitative picture.* Report 176. Sweden.

National Center for Education Statistics (2001). *Internet Access in US Public Schools and Classrooms: 1994-2000.* Washington, D.C. May.

OFTEL (1997a). *Access to the Internet for Schools. Consultation on BT's proposal.* July 1997.

OFTEL (1997b). *Access to the Superhighway for Schools: A statement following consultation on the regulatory framework for BT's prices for schools.* October 1997.

Policy Information Center (PIC) (1997). *Computers and Classrooms: The Status of Technology in U.S. Schools.* Policy Information Report. Princeton, NJ.

Prieger, J. (1998). Universal Service and the Telecommunications Act of 1996: The fact after the Act. *Telecommunications Policy*, 22,1.

SEC (2001) 222, *Benchmarking Report following-up the "Strategies for jobs in the Information Society.* ESDIS/Ministries of Education, February 2001.

Teligen (2000). *Report on Telecommunications Tariff Data.* Brussels: EC, DG INFSO.

U.S. Department of Education (USDE) (2000a). *E-Learning: Putting a world-class education at the fingertips of all children.* Washington, D.C.

U.S. Department of Education (USDE) (2000b). *E-Rate and the Digital Divide: A Preliminary Analysis from the Integrated Studies of Educational Technology*. Doc. # 00-17, prepared by Michael J. Puma, Duncan D Chaplin, and Andreas D. Pape, The Urban Institute, Washington, DC, 2000.

WIK (2000). *Study on the re-examination of the scope of universal service in the telecommunications sector of the EU, in the context of the 1999 Review*. Bad Honnef.